\documentclass[%
aip,
amsmath,amssymb,
preprint,%
reprint,%
]{revtex4-1}

\usepackage{graphicx}% Include figure files
\usepackage{dcolumn}% Align table columns on decimal point
\usepackage{bm}% bold math
\usepackage{lipsum}
\usepackage[dvipsnames]{xcolor}
\usepackage[utf8]{inputenc}
\usepackage[T1]{fontenc}
\usepackage{mathptmx}
\usepackage{etoolbox}
\usepackage{tabularx}
\usepackage{xcolor}
\usepackage{chemformula}
\usepackage{siunitx}
\usepackage{booktabs}

\newcommand{\blue}[1]{{\color{Black}{#1}}}
\makeatletter
\def\@email#1#2{%
	\endgroup
	\patchcmd{\titleblock@produce}
	{\frontmatter@RRAPformat}
	{\frontmatter@RRAPformat{\produce@RRAP{*#1\href{mailto:#2}{#2}}}\frontmatter@RRAPformat}
	{}{}
}%
\makeatother
\begin{document}
	\preprint{AIP/123-QED}
	
	\title{Lateral Solid Phase Epitaxy of Yttrium Iron Garnet}
	% Force line breaks with \\
	\author{Sebastian Sailler} \email{sebastian.sailler@uni-konstanz.de}
	\affiliation{Department of Physics, University of Konstanz, Konstanz, Germany}

	\author{Darius Pohl} \affiliation{Dresden Center for Nanoanalysis (DCN), Dresden, Center for Advancing Electronics Dresden (cfaed), TUD Dresden University of Technology, Dresden, Germany}
	
	\author{Heike Schlörb} \affiliation{Leibniz Institute for Solid State and Materials Research Dresden, Dresden, Germany}
	
	\author{Bernd Rellinghaus} \affiliation{Dresden Center for Nanoanalysis (DCN), Dresden, Center for Advancing Electronics Dresden (cfaed), TUD Dresden University of Technology, Dresden, Germany}
	
	\author{Andy Thomas} \affiliation{Leibniz Institute for Solid State and Materials Research Dresden, Dresden, Germany} \affiliation{TUD Dresden University of Technology, Dresden, Germany}
	
	\author{Sebastian T. B. Goennenwein} \affiliation{Department of Physics, University of Konstanz, Konstanz, Germany}
	
	\author{Michaela Lammel} \email{michaela.lammel@uni-konstanz.de}
	\affiliation{Department of Physics, University of Konstanz, Konstanz, Germany}

	\date{\today}% It is always \today, today,
	% but any date may be explicitly specified
	
	\begin{abstract}
		Solid phase epitaxy is a crystallization technique used to produce high quality thin films. Lateral solid phase epitaxy furthermore enables the realization of non-planar structures, which are interesting, e.g., in the field of spintronics. Here, we demonstrate lateral solid phase epitaxy of yttrium iron garnet over an artificial edge, such that the crystallization direction is perpendicular to the initial seed. We use single crystalline garnet seed substrates partially covered by a \ch{SiO_x} film to study the lateral crystallization over the \ch{SiO_x} mesa. The yttrium iron garnet layer retains the crystal orientation of the substrate not only when in direct contact with the substrate, but also across the edge on top of the \ch{SiO_x} mesa. By controlling the crystallization dynamics it is possible to almost completely suppress the formation of polycrystals and to enable epitaxial growth of single crystalline yttrium iron garnet on top of mesas made from \blue{ceramic} materials. From a series of annealing experiments, we extract an activation energy of \blue{\SI{3.0}{eV}} and a velocity prefactor of \blue{\SI{6.5e14}{nm/s}} for the lateral epitaxial crystallization along the <$100$> direction. Our results pave the way to engineer single crystalline non-planar yttrium iron garnet structures with controlled crystal orientation. 
	\end{abstract}
	
	\maketitle
	
	\section{Introduction}
	Epitaxy is one of the most commonly used techniques for obtaining single crystalline thin films. \cite{stringfellow_epitaxy_1982, evans_crystallization_2018} As a subset, solid phase epitaxy (SPE) describes the phase transition of an amorphous solid to its crystalline form while in contact with a crystalline seed of a similar or identical lattice parameter.\cite{johnson_solid-phase_2015} This causes the crystallization to start from the interface with the seed material and results in a single crystalline thin film of the same crystal orientation as the seed.\cite{johnson_solid-phase_2015}
	
	A special type of SPE is lateral solid phase epitaxy (LSPE), where the crystallization direction is perpendicular to the initial seed surface normal. \cite{ishiwara_lateral_1983, ishiwara_lateral_1986} Initially, lateral solid phase epitaxy was developed for the fabrication of silicon on insulator structures and has been an important technological step for the semiconductor industry. \cite{ishiwara_lateral_1983, ishiwara_lateral_1986, kusukawa_enhancement_1990, hoefflinger_itrs_2011} Therefore, the SPE of silicon and germanium has been studied most comprehensively. \cite{csepregi_regrowth_1976, csepregi_regrowth_1977_Ge, williams_solid_1983, johnson_intrinsic_2008_Ge} Recently, the lateral crystallization of oxide thin films has gained increasing interest and has been shown for \ch{Ba_{0.6}Sr_{0.4}TiO3},\cite{lee_introduction_2000} \ch{Nb:TiO2}\cite{taira_lateral_2014} and \ch{SrTiO3}. \cite{chen_seeded_2019}
	
	In this paper we investigate the oxide compound yttrium iron garnet (\ch{Y3Fe5O12}, YIG). Its ferrimagnetic properties, \cite{althammer_quantitative_2013} combined with a long spin diffusion length \cite{kajiwara_transmission_2010, cornelissen2015long} as well as an exceptionally low Gilbert damping and a low coercive field \cite{chang_nanometer-thick_2014, hauser_yttrium_2016} make it a prototypical material in the field of magnetism and spintronics. \cite{cherepanov_saga_1993} 
	
	\blue{For typical spintronic experiments, high quality YIG thin films are needed, as the magnetic and structural properties are connected.\cite{haidar_thickness-_2015, chang_nanometer-thick_2014} Over the last approximately 50 years, a multitude of publications for a variety of deposition techniques have shown that getting single crystalline YIG thin films with excellent quality on lattice matched substrates such as gadolinium gallium garnet is indeed possible.\cite{beaulieu_temperature_2018, dubs_low_2020, krockenberger_solid_2008, hauser_yttrium_2016, jang_new_2001, chang_nanometer-thick_2014, Lammel_ALD_PhysRevMaterials} Only recently the exact crystallization dynamics from the amorphous phase on these substrates has been reported.\cite{sailler_crystallization_2023} However, up to now the literature almost exclusively describes the vertical crystallization (the crystallization along the film surface normal) of YIG thin films on a fully coated substrate\cite{beaulieu_temperature_2018, dubs_low_2020, krockenberger_solid_2008, hauser_yttrium_2016, jang_new_2001, chang_nanometer-thick_2014, Lammel_ALD_PhysRevMaterials} and the few existing reports using a different approach do not give details concerning the dynamics of the lateral crystallization.\cite{heyroth_monocrystalline_2019}$^{,}$\footnote{Heyroth et al. \cite{heyroth_monocrystalline_2019} showcased a single crystalline YIG bridge fabricated by coating a resist template with YIG deposited via pulsed laser deposition. However, we do not expect their annealing process to be easily transferrable to larger, sputtered YIG structures, as we see significant formation of polycrystalline YIG above \SI{660}{\degreeCelsius}.}
	
	In this work, we quantitatively analyze the lateral solid phase epitaxy of YIG over an artificial mesa on top of crystalline seed substrates. From systematic annealing experiments we extract the activation energy as well as the crystallization velocity which allow for a full description of the lateral solid phase epitaxy. Apart from being interesting from a material science point of view, the lateral approach described here opens two interesting avenues. First, it is a key step towards the development of free standing and easily detachable single crystalline YIG layers, which can be used in experiments that are otherwise impeded by the substrate materials.\cite{kosen_microwave_2019, barman_2021_2021, guo_low_2022} Secondly, the possibility to realize epitaxial non-planar YIG layers facilitates novel experimental solutions to implement for example non-planar magnon transport and wave computing interconnects.\cite{chumak_magnon_2015, sakharov_spin_2020, barman_2021_2021, salazar-cardona_nonreciprocity_2021, chumak_advances_2022}}
	
	%In these areas the focus in recent years was expanded towards non-planar, three dimensional and curved magnetic structures \cite{sanz-hernandez_fabrication_2017, fernandez-pacheco_three-dimensional_2017, heyroth_monocrystalline_2019, Lammel_ALD_PhysRevMaterials} as curvature was reported to induce novel phenomena like curvature-induced anisotropy \cite{Streubel_2016} or the Dzyaloshinskii-Moriya-interaction. \cite{dzyaloshinsky_thermodynamic_1958} These phenomena in turn are predicted to lead to a variety of resulting effects, for example to spin-wave nonreciprocities \cite{otalora_curvature-induced_2016, salazar-cardona_nonreciprocity_2021} or magnetochiral effects.\cite{dietrich_influence_2008, sloika_curvature_2014} 
	
	%Realizing non-planar magnetic structures is therefore highly desirable. However, the deposition techniques commonly used for the fabrication of YIG like pulsed laser deposition and magnetron sputtering typically yield planar thin films.
	
 	%In this work, we report the lateral solid phase epitaxy of YIG over an artificial mesa on top of crystalline seed substrates. From systematic annealing experiments, we extract the activation energy as well as the crystallization velocity which allow for a full description of the lateral solid phase epitaxy. Our results pave the way for experiments on more sophisticated non-planar structures of one of the prototypical magnetic materials for spintronics.
		
	\section{Methods}
	 All films discussed in this publication were deposited using radio-frequency magnetron sputtering at room temperature in an AJA International sputtering system.
	
	For the lateral crystallization experiments, we used yttrium aluminum garnet (\ch{Y3Al5O12}, YAG, \textit{CrysTec}) substrates with the <$111$> crystal orientation being parallel to the surface normal as well as two types of gadolinium gallium garnet (\ch{Gd3Ga5O12}, GGG, \textit{SurfaceNet}) substrates, where the crystal orientation along the surface normal is either <$111$> or <$001$>. Since GGG and YAG crystallize in the same space group Ia$\bar{3}$d as YIG and their lattice parameters are comparable to those of YIG (a$_{\mathrm{GGG}}$ = \SI{1.2376}{nm},\cite{Gates_Rector_2019_GGG} a$_{\mathrm{YAG}}$ = \SI{1.2009}{nm},\cite{Gates_Rector_2019_YAG} a$_{\mathrm{YIG}}$ = \SI{1.2380}{nm}\cite{Gates_Rector_2019_YIG}) they are considered as closely lattice matched. Before the sputtering process all substrates were cleaned for five minutes in aceton and isopropanol, and one minute in deionized water in an ultrasonic bath. 
	
\begin{figure}[t]
		\begin{center}
			\includegraphics[width=\linewidth]{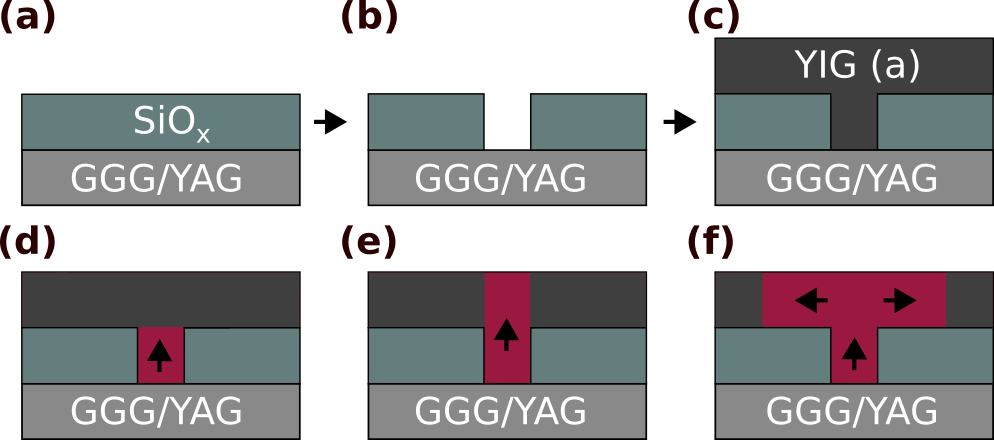}
			\caption{Sample preparation scheme for lateral crystallization. \blue{After coating the single crystalline substrate with \ch{SiO_{x}}(a), a 1mm wide trench in the \ch{SiO_x} is defined by optical lithography and subsequent etching (b)}. YIG is afterwards sputtered on top (c) and annealed at \SI{600}{\degreeCelsius} - \SI{650}{\degreeCelsius}, which induces the crystallization at the substrate interface (d). After vertically crystallizing (e) the crystallization front propagates laterally, over the edge of the \ch{SiO_x} mesa (f).}
			\label{Fig_1_lat}
		\end{center}
	\end{figure}

	To create an artificial mesa on the substrate surface, we first sputter a nominally \SI{20}{nm} thick \ch{SiO_{x}} layer either from a \ch{SiO2} sinter target or by reactive sputtering from a silicon target onto one of the garnet substrates (cp. Fig.~\ref{Fig_1_lat}(a)). The \ch{SiO_x} layer from the \ch{SiO_2} target (reactive Si) was deposited at a sputtering pressure of \SI{2.7e-3}{mbar} in pure argon atmosphere (argon to oxygen 13:4) at \SI{150}{W} (\SI{100}{W}) at a rate of \SI{0.0208}{nm/s} (\SI{0.0172}{nm/s}).
	
	Several ways were utilized to fabricate the mesa from the \ch{SiO_x} layer (cp. Fig.~\ref{Fig_1_lat}(b)). Some of the substrates were partially covered with Kapton tape before the \ch{SiO_x} sputtering process, which yield the desired structure by simply removing the tape afterwards. While this path is simple, it leads to comparably rough edges. To improve the edge quality of the \ch{SiO_x} mesa optical lithography and subsequent etching of the \ch{SiO_{x}} was utilized. The form of the mesa was first defined in photoresist by a Smart Print (\textit{Smartforce Technologies}) and after developing transferred into the \ch{SiO_x} by physical etching with a \ch{SF6} plasma in an Oxford Instrument RIE system. To counteract any potential damage to the substrate after the etching step, it was annealed at \SI{700}{\degreeCelsius} for \SI{4}{h}. We also used chemical etching of the \ch{SiO_{x}} stripe in buffered \ch{HF}.
		
	In a subsequent fabrication step, we then deposit the YIG film on top of the predefined \ch{SiO_x} mesas as sketched in Fig.~\ref{Fig_1_lat}(c). As we want to investigate lateral crystallization, the nominal YIG thickness was chosen to be at least twice as thick as the \ch{SiO_x} to ensure continuity of the YIG layer across the mesa edge. YIG was sputtered from a stochiometric sinter target at \SI{2.7e-3}{mbar} argon pressure and \SI{80}{W} power, at a rate of \SI{0.0135}{nm/s}. This results in a complete coverage of the \ch{SiO_x} mesas with YIG, where the YIG film within the predefined trench is still in contact with the lattice matched substrate (cp. Fig.~\ref{Fig_1_lat}(c)).
	
	To induce and observe crystallization of the YIG layer, the complete stack was then annealed at temperatures between \SI{600}{\degreeCelsius} and \SI{650}{\degreeCelsius} multiple times in a tube furnace under air. The expected crystallization behavior upon annealing is shown in Fig.~\ref{Fig_1_lat}(d-f). First, YIG starts crystallizing vertically from the lattice matched substrate via solid phase epitaxy (cp. Fig.~\ref{Fig_1_lat}(d)). After reaching the top edge of the film, the epitaxial, single crystalline YIG now acts as a seed for the amorphous YIG on \ch{SiO_x} (cp. Fig.~\ref{Fig_1_lat}(e)). Starting from the edge of the mesa a lateral crystallization front is expected to move with a constant velocity (cp. Fig.~\ref{Fig_1_lat}(f)). The evolution of crystalline YIG across the mesa was observed in a scanning electron microscope (SEM, \textit{Zeiss GeminiSEM}), where the crystalline region was analyzed via electron backscatter diffraction (EBSD). 
	
	Transmission electron microscopy was conducted using a JEOL JEM F200 operated at \SI{200}{kV} acceleration voltage equipped with a GATAN OneView CMOS camera for fast imaging. Local EDS analysis was perfomed using a dual \SI{100}{m\square m} window-less silicon drift detector.
	
	\section{Results and Discussion}
	The selection of the ideal annealing temperature is crucial for the observation of lateral crystallization. In our previous work\cite{sailler_crystallization_2023} we determined the parameters describing the vertical crystallization of YIG for different time and temperature pairs depending on the substrate. We demonstrated, that epitaxial crystallization from a lattice matched seed becomes possible at temperatures below those required for the formation of polycrystals. Vice-versa, the formation of polycrystalline YIG can be suppressed by using sufficiently low annealing temperatures. From our results we approximated, that for $T < \SI{660}{\degreeCelsius}$ the formation of a fully polycrystalline film on \ch{SiO_x} would take about \SI{100}{h}. Avoiding the formation of polycrystalline grains is of great importance, as those would hinder the epitaxial crystallization. Please note that nucleation is a thermally activated process and therefore statistically also possible for lower temperatures. During our study we found that nucleation was more likely to occur if there were external nucleation sites in the form of particles on the respective sample, demonstrating the need of clean surfaces. However, the temperature for annealing has to be sufficiently high, since the crystallization rate depends exponentially on the temperature (cp. Eq.~\eqref{eq_v_arr}). Below \SI{540}{\degreeCelsius} the crystallization via solid phase epitaxy of YIG will not occur in reasonable times (t >\SI{100}{h}). Therefore, we are confined to a temperature range of about \SI{120}{\degreeCelsius} (\SI{540}{\degreeCelsius} - \SI{660}{\degreeCelsius}) to study the lateral crystallization of YIG.
	
	\begin{figure*}[t!]
		\begin{center}
			\includegraphics[width=\linewidth]{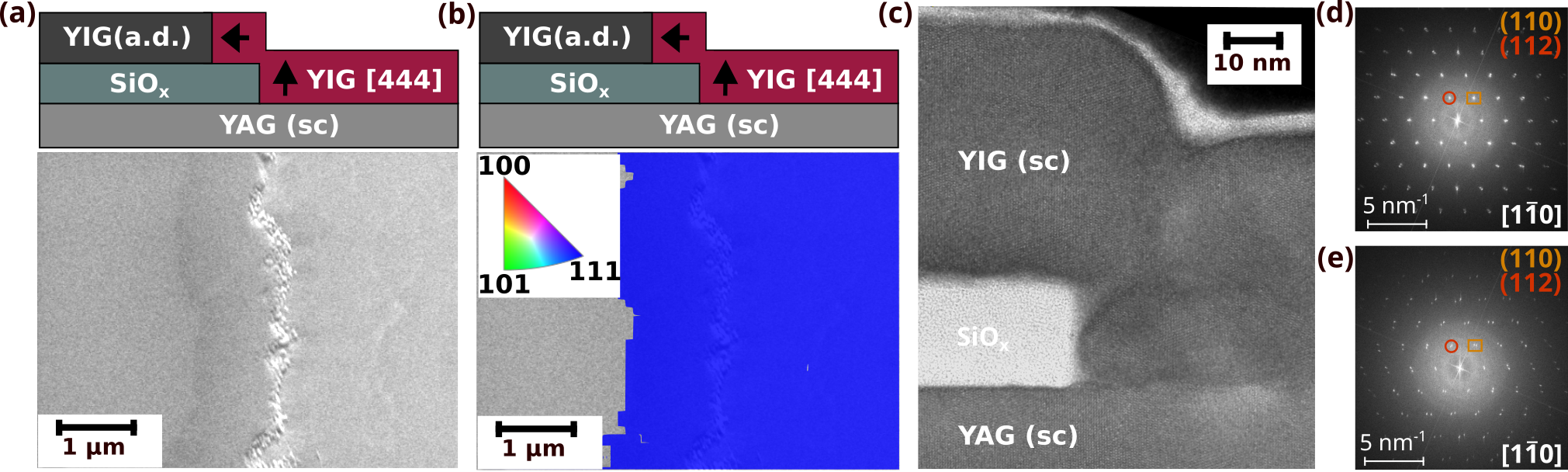}
			\caption{Observation of lateral crystallization of YIG on a patterned YAG/\ch{SiO_x} substrate after annealing for \SI{96}{h} at \SI{600}{\degreeCelsius}. (a) Secondary electron (SE) image of the mesa etched into the \ch{SiO_{x}}, where the left side is elevated as sketched. First, the YIG layer on top of the YAG substrate crystallizes vertically towards the top of the sample and thereby changes from the as deposited state (a.d.) into a single crystalline (sc) YIG (right side of the image). Once the YIG reaches the top edge of the sample, it crystallizes laterally via LSPE from right to left onto the \ch{SiO_x} layer. The formation of crystalline YIG is accompanied by a change in contrast. \blue{Superimposing the SE image from (a) with the EBSD mapping (b) confirms the single crystalline nature of the YIG layer also in the lateral crystallization regime. The color code is taken from the inverse pole figure in the out of plane direction and confirms the epitaxial growth in the crystal direction given by the substrate.} The TEM investigation in (c) further verifies the lateral solid phase epitaxy of YIG on top of the \ch{SiO_x}. Directly next to the sharp edge of the \ch{SiO_x} mesa a slight rotation in the YIG crystal orientation can be seen, which transfers to the single crystalline YIG on top of the \ch{SiO_x}, as it grows epitaxially from the YIG seed. \blue{To visualize this rotation, the fast Fourier transforms (FFT) of two TEM images of YIG crystallized laterally on a \ch{SiO_x} mesa, one near the mesa edge and one near the end of the crystallization front, are depicted in (d) and (e), respectively. Here, the zone axis is along $[1\overline{1}0]$ and two reflections along $(110)$ and $(112)$ are marked.}}
			\label{Fig_2_lat}
		\end{center}
	\end{figure*}
	
	\blue{To mathematically describe the lateral solid phase epitaxy of YIG from a YIG seed we use a modified Arrhenius equation, which is commonly used for the description of homoepitaxial crystallization out of the amorphous phase.\cite{csepregi_chaneling_1975, white_ion_1988, johnson_intrinsic_2008_Ge,johnson_solid-phase_2015}} This description assumes a homogeneous crystallization front that moves through an amorphous material starting from a crystalline seed of the same material. At a given temperature $T$, this crystallization front is expected to move with a constant velocity ($v = \frac{l}{t}$). Here $v$ is the lateral crystallization velocity, $l$ the lateral crystallization distance and $t$ the annealing time. The crystallization velocity itself depends exponentially on the temperature as well as the activation energy $E_A$ and can be described by Eq. \eqref{eq_v_arr}, \cite{csepregi_chaneling_1975, johnson_solid-phase_2015}
	\begin{equation}
		v = v_0 \cdot e^{\frac{-E_A}{k_B\cdot T}}
		\label{eq_v_arr}
	\end{equation}
	where $k_B$ is the Boltzmann constant and the prefactor $v_0$ represents a maximal velocity. 

	To confirm LSPE of YIG and to determine the crystallization velocities $v(T)$, we analyze our samples with electron back scatter diffraction (EBSD) for a sequence of annealing times. EBSD also allows us to exclude polycrystalline YIG on top of the \ch{SiO_x} mesa. A secondary electron (SE) image taken across the edge of the mesa is depicted in Fig.~\ref{Fig_2_lat}(a). On the right, crystalline YIG is on top of the YAG substrate as depicted in the schematics above the SE image. The lateral crystallization front moves from the right across the mesa's edge, which can be seen in the middle of the image, towards the left. The area of crystallized YIG on top of the \ch{SiO_x} mesa can be discerned as a change in gray level which we ascribe to the height and density change upon crystallization. 
	
	To verify the formation of single crystalline YIG on YAG as well as across the mesa on \ch{SiO_x}, the SE image is superimposed with the results from the EBSD mapping (cp. Fig.~\ref{Fig_2_lat}(b)). \blue{The monochrome color of the inverse pole figure from the out of plane direction confirms an epitaxial formation of single crystalline YIG initiated by the substrate.} This demonstrates that we achieved lateral solid phase epitaxy of \SI{1}{\micro \meter} YIG over a \SI{18}{nm} high mesa. Furthermore, unlike in similar studies on Si and other oxides, \cite{ishiwara_lateral_1983, ishiwara_lateral_1986, chen_seeded_2019, taira_lateral_2014} we do not find any polycrystalline YIG seeds on \ch{SiO_{x}} after \SI{96}{h} at \SI{600}{\degreeCelsius}.
	
	To further investigate the single crystalline nature of the laterally crystallized YIG, transmission electron microscopy (TEM) was performed. Fig.~\ref{Fig_2_lat}(c) depicts a side view of the mesa structure as illustrated schematically in Fig.~\ref{Fig_1_lat}(f) and Fig.~\ref{Fig_2_lat}(a + b). The TEM images show a sharp edge of the deposited \ch{SiO_x} and a rounder YIG edge on top (Fig.~\ref{Fig_2_lat}(c)). \blue{On top of the YIG a conducting layer consisting of Au/Pd/Pt was utilized, which prevents an amorphization of the otherwise isolating sample by the ion beam due to charging effects.} The TEM image supports the results of the SEM investigation, in that the YIG layer crystallizes epitaxially from the substrate and also laterally on top of the \ch{SiO_x} layer. However, close to the \ch{SiO_x} edge in Fig.~\ref{Fig_2_lat}(c), some imperfections in the single crystal can be resolved in the TEM image. \blue{Additionally, a rotation of the crystal is visible. A tilted YIG crystallizing from the bottom acts as the new seed for the lateral crystallization and therefore causes the YIG crystal on top of the \ch{SiO_x} to be slightly rotated with respect to the substrate. As this rotation takes place over an extended distance, it is not prevalent in Fig.\ref{Fig_2_lat}(c). To visualize the rotation we include the fast Fourier transforms (FFT) obtained from two TEM images: one near the mesa edge (cp. Fig.~\ref{Fig_2_lat}(d)) and one at the end of the crystallization front (cp. Fig.~\ref{Fig_2_lat}(e)). In the FFTs, a clear rotation of the YIG peaks is observed. This rotation is visible in samples where the surface normal is parallel to the $[111]$ direction, but not in samples with the surface normal direction parallel to $[100]$. Additionally, the rotation takes place exclusively in planes perpendicular to the $[1\overline{1}0]$ direction of YIG.}
		
	The distance up to which the YIG crystallizes laterally on top of the \ch{SiO_x} can be extracted from both techniques, TEM and SEM. As the SEM allows for a fast measurement of the crystallization front, the lateral crystallization data is extracted from the SE images and EBSD data. 
		
	\blue{To determine the lateral crystallization velocities, each sample is annealed multiple times and the distance covered by the crystallization front is measured after each annealing step. At each time and temperature pair we analyze multiple images across the mesa edge and thereby obtain the respective, average crystallization distance and its standard deviation. The time which corresponds to a lateral crystallization distance of \SI{0}{nm} is approximated by the time necessary for the YIG to crystallize vertically up to the height of the step, e.g., \SI{0.34}{h} for a step height of \SI{20}{nm} with a vertical crystallization velocity of \SI{58.8}{nm/h} for YIG on GGG at \SI{600}{\degreeCelsius}, as detailed in our previous work.\cite{sailler_crystallization_2023} We then determine the lateral crystallization velocity from the slope of a linear fit to the data.}
		
	\begin{figure}[t]
		\begin{center}
			\includegraphics[width=\linewidth]{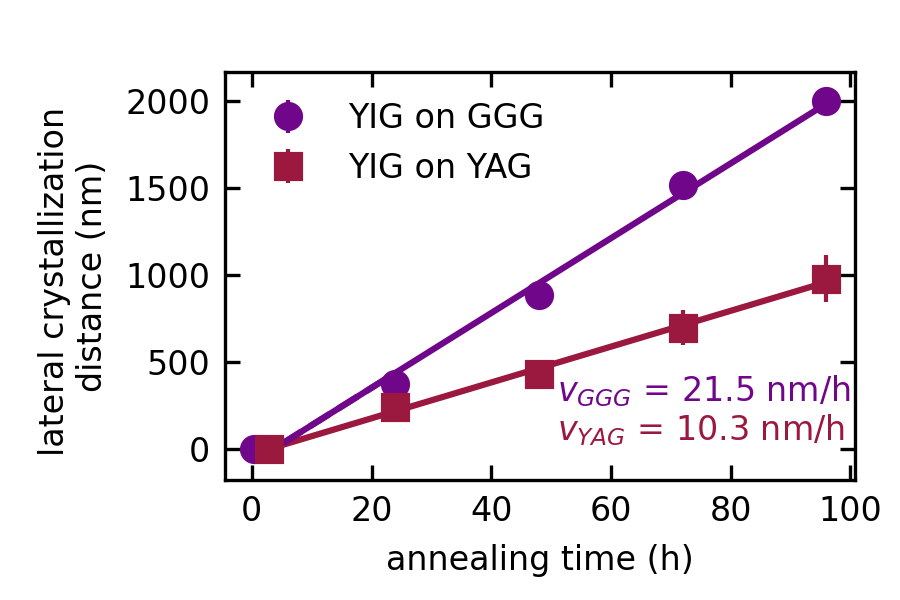}
			\caption{Lateral crystallization velocities for YIG at \SI{600}{\degreeCelsius} on two different seed substrates. After annealing for \SI{24}{h} the lateral crystallization distance was extracted from multiple SEM images of different areas over the mesa edge and the sample annealed again. The average lateral crystallization with the standard deviation is depicted over the annealing time. A linear fit to the data was used to extract the lateral crystallization rate and the delay before the onset of lateral crystallization. While YIG starts to grow earlier on YAG it shows a slower rate of \SI{10.3}{nm/h} compared to the \blue{\SI{21.5}{nm/h}} on GGG.}
			\label{Fig_3_lat}
		\end{center}
	\end{figure}
	
	Figure~\ref{Fig_3_lat} shows the lateral crystallization velocity of YIG at \SI{600}{\degreeCelsius} when using <$111$> oriented YAG and GGG as seed substrates. We extract a lateral crystallization velocity of $v_{\mathrm{YAG}}$ = \SI{10.3}{nm/h} (\SI{0.003}{nm/s}) for YIG on YAG and $v_{\mathrm{GGG}}$ = \blue{\SI{21.5}{nm/h} (\SI{0.006}{nm/s})} for YIG on GGG.
	
	Since the lateral crystallization starts from a YIG seed for either of the substrates, one might expect the same lateral crystallization velocity. However, the different velocities suggest that the substrate indeed influences the maximal crystallization velocity. On the one hand, this behavior might originate from a different crystalline quality of the vertically crystallized YIG on the two substrates. Since GGG exhibits a lower lattice mismatch than YAG, it is expected to lead to a higher quality YIG film by epitaxy. On the other hand, the crystallization was not perfectly epitaxial near the \ch{SiO_x} mesa (cp. Fig.~\ref{Fig_2_lat}(c)), which could influence the initial crystallization as well as the final velocity. In the course of this work we observed that the final velocity depends on surface and mesa edge quality. 
	
	To further substantiate our results, we compare our lateral crystallization velocity to the vertical crystallization velocity of YIG, which we reported in earlier work.\cite{sailler_crystallization_2023} \blue{Compared to the vertical crystallization velocity of \SI{58.8}{nm/h} (\SI{0.016}{nm/s}) for a YIG thin film on GGG, the lateral crystallization velocity of YIG on GGG $v_{\mathrm{GGG}}$ is 3.6 times slower. A similar trend is seen on YAG, where the lateral crystallization velocity $v_{\mathrm{YAG}}$ is roughly 1.6 times slower than the vertical crystallization velocity of YIG on YAG of \SI{16.2}{nm/h} (\SI{0.0045}{nm/s}). We speculate that the YIG crystal quality impacts the crystallization speed, but more detailed experiments will necessary to resolve this point in the future.} Since the vertical crystallization of YIG on GGG was confirmed to be epitaxial, it allows for a good comparison with the LSPE of YIG. The observation that the lateral crystallization is slower than the vertical one, is na\"ively unexpected but was also reported for silicon, where the lateral crystallization was four to eight times slower compared to the vertical direction.\cite{ishiwara_lateral_1983} In silicon this behavior was ascribed back to the formation of facets and defects in the lateral silicon. 
	
	Another reason for the slower lateral crystallization velocity could also be a dependence on the crystal direction along which the YIG crystallizes. From studies on silicon it is known, that differences in vertical crystallization velocity \cite{csepregi_regrowth_1976, csepregi_regrowth_1977_Ge} are transferred into the lateral crystallization. \cite{kusukawa_enhancement_1990} Such a crystal orientation dependence of the crystallization velocities has also been reported for bulk YIG. \cite{nielsen_growth_1958, erk_effect_1980, tolksdorf_facet_1981}

	Hence, to compare the lateral crystallization velocities extracted here with the ones from our previous work, that describe the vertical crystallization, the direction along which the crystallization takes place needs to be taken into account.\footnote{The direction of the lateral crystallization of the YIG layers from Fig.~\ref{Fig_3_lat} can be identified to be $[11\overline{2}]$, as it is perpendicular to $[111]$ and $[1\overline{1}0]$ known from the substrate. There, the $[111]$ direction is parallel to the surface normal of the substrate, which is cut in a way to have one edge aligned parallel to the $[1\overline{1}0]$ direction. The vertical crystallization on GGG took place along the $[111]$ direction.\cite{sailler_crystallization_2023}} Together with the different seed substrates, this direction dependence could play a role in the difference between the lateral crystallization velocity and the vertical one of a factor two for YIG on GGG.
	
	In addition to the possibility of a direction dependence the lateral solid phase epitaxy of YIG is expected to exponentially depend on the temperature as described in Eq. \eqref{eq_v_arr}.
	
	To investigate both the crystal orientation dependence and the temperature dependence, multiple samples were prepared on GGG substrates with the $[001]$ and $[111]$ directions parallel to the surface normal. These orientations allow for the investigation of the lateral crystallization velocity along the $[010]$, $[100]$, $[1\overline{1}0]$ and $[11\overline{2}]$ directions. As YIG is a cubic system, these will be referred to as their equivalents <$100$>, <$110$> and <$112$>. The crystallization along these directions was evaluated for samples annealed at different temperatures of \SI{600}{\degreeCelsius}, \SI{625}{\degreeCelsius}, \SI{637}{\degreeCelsius} and \SI{650}{\degreeCelsius}. 
	
	Although the formation of some polycrystalline grains could be seen with increasing temperature, this did not hinder the lateral crystallization up to a distance of \SI{2.2}{\micro \meter} via LSPE. \blue{This distance results from our choice of maximum annealing time at each tested temperature. Please note that while in principle the lateral crystallization should continue through the whole layer, for very long times and/or higher annealing temperatures the statistical formation of grains can impede the propagation of the crystallization front.}
	
	\begin{table}[b]
		\centering
		\newcolumntype{Y}{>{\centering\arraybackslash}X}
		\caption{Activation energies $E_A$ and prefactors $v_0$ of the epitaxial YIG from our work in comparison with silicon and germanium and \ch{SrTiO3}.}
		\begin{tabularx}{\linewidth}{ Y Y Y  Y Y  }
			\hline \hline
			Material & Ref & Orientation &$E_A$ (eV) & $v_0$ ($\mathrm{nm}/\,\mathrm{s}$) \\ \toprule
			YIG & this work & <$100$> & $3.0\,\pm\,0.2$ & {\SI{6.5e14}{}} \\
			YIG & this work & <$110$> & $3.0\,\pm\,0.2$ & {\SI{6.5e14}{}}\\
			YIG & this work & <$112$> & $3.0\,\pm\,0.2$ & {\SI{6.9e14}{}} \\ \midrule
			Si & Csepregi\cite{csepregi_chaneling_1975} & <$100$> & $2.3\,\pm\,0.1$ & \SI{1.5e13}{} \\ 
			Si & Csepregi\cite{csepregi_chaneling_1975} & <$110$> & $2.3\,\pm\,0.1$ & \SI{6.4e12}{} \\ \midrule
			Ge  & Johnson\cite{johnson_intrinsic_2008_Ge} & <$100$> & $2.15\,\pm\,0.04$ & \SI{2.6e16}{} \\ \midrule
			\ch{SrTiO3} & White\cite{white_ion_1988} & <$100$> & 0.77 & $5\times 10^5$\\
			\ch{SrTiO3} & Chen\cite{chen_distinct_2017} & <$100$> & 0.7 & $7.8\times 10^2$\\ \hline \hline \label{tab_1}
		\end{tabularx}
	\end{table}
	
\begin{figure}[t]
		\begin{center}
			\includegraphics[width=\linewidth]{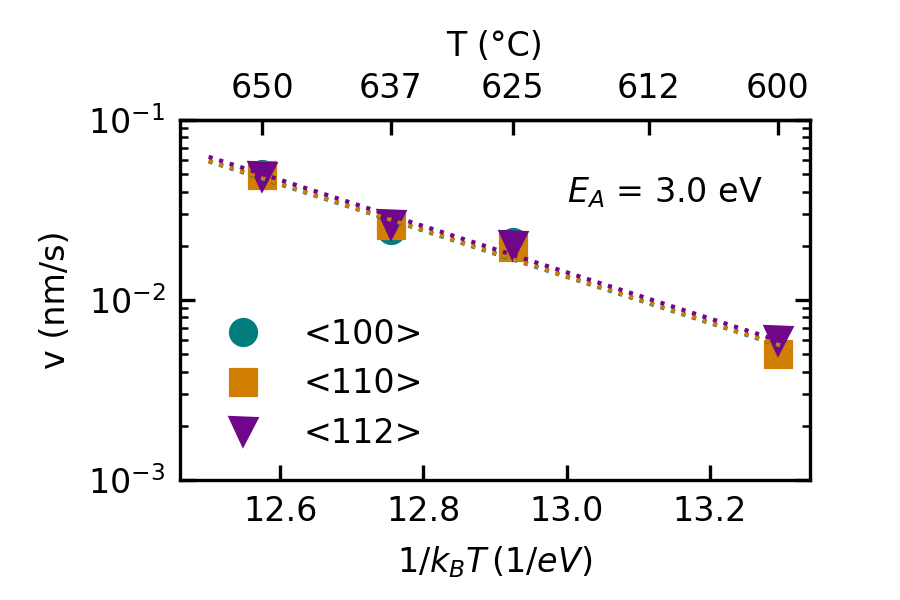}
			\caption{Crystallization velocities as a function of temperature along different crystal directions on a GGG substrate. A semi-logarithmic plot over the inverse temperature yields the activation energy $E_A$ as well as the maximal rates $v_0$. For YIG an activation energy of $E_A$ = \blue{\SI{3.0}{eV}} is found. The crystallization velocity depends only marginally on the direction and is $v_0 $(<$100$>) = \blue{\SI{6.5e14}{nm/s}}, $v_0 $(<$110$>) = \blue{\SI{6.5e14}{nm/s}} and $v_0 $(<$112$>) = \blue{\SI{6.9e14}{nm/s}}.}
			\label{Fig_4_lat}
		\end{center}
	\end{figure}

	Fig.~\ref{Fig_4_lat} shows the results of these series over the annealing temperature. Each velocity shown in Fig.~\ref{Fig_4_lat} is extracted from a series like the one shown in Fig.~\ref{Fig_3_lat}. From the semi-logarithmic plot, a linear dependence of the lateral crystallization velocity on the inverse temperature can be seen. Here, the lateral crystallization velocity increases from  \SI{21.5}{nm/h} (\SI{0.006}{nm/s}) at \SI{600}{\degreeCelsius} up to \SI{173.1}{nm/h} (\SI{0.048}{nm/s}) at \SI{650}{\degreeCelsius}. As all the samples are made of the same material, i.e. YIG, we expect one single activation energy for all directions, \cite{csepregi_chaneling_1975, csepregi_regrowth_1977_Ge, williams_solid_1983, johnson_intrinsic_2008_Ge, claverie_amorphization_2010} which we extract from the slope of a linear fit to all velocities. This results in an activation energy of $E_A$ = 3.0 $\pm$ \SI{0.2}{eV} for the lateral crystallization of YIG. 
	
	In contrast to the reports for the formation of bulk YIG,\cite{tolksdorf_facet_1981} however, we find no significant difference in the maximal lateral crystallization velocity depending on crystal orientation. For the formation of bulk YIG from the liquid phase it was reported, that facets in <$110$> and <$112$> direction are the thermodynamically most stable, while the <$111$> direction was described to grow fastest. \cite{erk_effect_1980, tolksdorf_facet_1981, beregi_dissolution_1983} There, YIG was found to crystallize up to 10 times faster along the <$111$> than along the <$110$> direction, \cite{erk_effect_1980} while the crystallization velocities along the <$110$> and <$112$> directions were found to behave very similarly. \cite{tolksdorf_facet_1981}
	
	For the LSPE of our sputtered thin films with an activation energy of \blue{\SI{3.0}{eV}} we find prefactors of $v_0 $(<$100$>) = \blue{\SI{6.5e14}{nm/s}}, $v_0 $(<$110$>) = \blue{\SI{6.5e14}{nm/s}} and $v_0 $(<$112$>) = \blue{\SI{6.9e14}{nm/s}}. Tolksdorf et al. reported a very similar growth behavior \blue{of YIG grown by liquid phase epitaxy} for facets along the <$110$> and <$112$> direction, with the <$112$> direction being slightly faster, which we find here as well.\cite{tolksdorf_facet_1981} No qualitative literature data could be found for the crystallization along the <$100$> direction. Further studies involving a lateral growth along the faster crystallizing <$111$> direction \cite{erk_effect_1980, tolksdorf_facet_1981} could help to verify a orientation dependence of lateral YIG growth. 
		
	Both the activation energies $E_A$ and the prefactors $v_0$ are in good agreement with the literature for solid phase epitaxy, see Tab.~\ref{tab_1}. Compared to the model systems of silicon, germanium and \ch{SrTiO3}, the activation energy for YIG is higher, while the crystallization velocities are in a similar order of magnitude as for silicon and germanium. The epitaxial crystallization process of YIG seems to be more similar to elemental Si and Ge than to the oxide \ch{SrTiO3}. 
	
	Additionally, our activation energy of $E_A$ = \blue{\SI{3.0}{eV} $\pm$ \SI{0.2}{eV}} for epitaxial YIG compares well with previously reported values. Specifically, investigations of YIG thin films on GGG revealed an activation energy of \blue{\SI{3.98}{eV}}.\cite{sailler_crystallization_2023} \blue{This increased activation energy on a not perfectly lattice matched substrate can be seen as the additional barrier needed to be overcome for starting the crystallization process.} For the formation of bulk, polycrystalline YIG from oxide powders, a value of \SI{5.08}{eV} was reported. \cite{wan_ali_investigation_2016} Chen et al.\cite{chen_distinct_2017} report, that the activation energy for epitaxial \ch{SrTiO3} is half of that of polycrystalline \ch{SrTiO3}, which is in good agreement with our findings for the YIG thin films. The activation energy for solid phase epitaxy of $E_A$\,=\,\blue{\SI{3.0}{eV}} is also roughly half of \SI{5.08}{eV} for the oxide powders and also reduced compared to the value of vertical crystallization on GGG substrates. We therefore conclude, that the lateral solid phase epitaxy of YIG is described by an activation energy of $E_A$ = \blue{\SI{3.0}{eV}} and for the directions <$100$>, <$110$> and <$112$> by the $v_0$ values of \blue{\SI{6.5e14}{nm/s}}, \blue{\SI{6.5e14}{nm/s}} and \blue{\SI{6.9e14}{nm/s}}, respectively.
		
	\section{Conclusion}
	To assess the lateral solid phase epitaxy of YIG, we defined \ch{SiO_x} mesa structures on top of single crystalline garnet substrates, which were subsequently covered by an amorphous YIG layer by room temperature sputtering. By carefully choosing the annealing temperature we were able to laterally crystallize up to \SI{2.2}{\micro \meter} of single crystalline YIG on top of an amorphous \ch{SiO_x} layer. At \SI{600}{\degreeCelsius} on GGG a crystallization velocity of \blue{\SI{21.5}{nm/h} (\SI{0.006}{nm/s})} was found, which increased by a factor seven to \SI{173.1}{nm/h} (\SI{0.048}{nm/s}) at \SI{650}{\degreeCelsius}. By extracting multiple lateral crystallization velocities at different temperatures and along different crystal orientations, we confirmed an exponential dependence on temperature as expected for LSPE. The resulting crystallization parameters are summarized in Tab.~\ref{tab_1}, where the crystallization velocity we derive is mostly independent on the crystal orientation of the seed. \blue{The understanding of these dynamics allows for a controlled and precise manufacturing of single crystalline YIG thin films of micrometer length scales on various substrates.}
	
	\section{Acknowledgments}
	This work was funded by the Deutsche Forschungsgemeinschaft (DFG, German Research Foundation) – Project-ID 446571927 and via the SFB 1432 - Project-ID 425217212. We gratefully acknowledge technical support and advice by the nano.lab facility of the University Konstanz. We acknowledge the use of the facilities in the Dresden Center for Nanoanalysis (DCN) at the Technische Universität Dresden and the support of Alexander Tahn. 
	
	\section{References}
	\bibliographystyle{apsrev4-1}
	\bibliography{PaperBib_lat.bib}% Produces the bibliography via BibTeX.
	
\end{document}